\magnification=\magstep1
\tolerance 500
\rightline{TAUP 2808/05}
\rightline{24 August, 2005}
\rightline{Revision 10 December, 2005}
\bigskip
\centerline{\bf On the Significance of a Recent Experiment}
\centerline{\bf Demonstrating  Quantum Interference in Time}
\bigskip
\centerline{Lawrence P. Horwitz\footnote{*}{e-mail:larry@post.tau.ac.il.}}
\centerline{School of Physics, University of Tel Aviv, Ramat Aviv
69978, Israel}
\centerline{and}
\centerline{Department of Physics, College of Judea and Samaria, Ariel, Israel}
\bigskip
\noindent{\it Abstract}
\par I discuss the interpretation of a recent experiment showing
quantum interference in time.  It is pointed out that the standard
nonrelativistic quantum theory, used by the authors in their analysis,
does not have the property of coherence in time, and hence
cannot account for the results found. Therefore, this
experiment has fundamental importance beyond the technical advances it
represents.  Some theoretical structures which consider the time as an
observable, and thus could, in principle, have the required coherence
in time, are discussed briefly, and the application of Floquet theory and the
manifestly covariant quantum theory of Stueckelberg are treated in
some detail.  In particular, the latter is shown to account for the
results in a simple and consistent way.
\bigskip 
\par The recent experiment of Lindner, {\it et al}$^1$, clearly shows the
effect of quantum interference in time for the  wave function of a
 particle.  The results are discussed in that paper in terms of a very precise
solution of the time-dependent nonrelativistic Schr\"odinger
equation. In this note, I wish to point out that the
nonrelativistic Schr\"odinger theory cannot be used to predict interference
phenomena in time, and therefore the very striking results of this
beautiful experiment have a fundamental importance which goes beyond
the technical advances which they represent. They imply, in fact, that
 the time variable $t$ must be adjoined to the set of standard quantum
 variables so that the standard ket $|x,t>$ for the representation of
 the quantum state (in Dirac's terminology$^2$) can be constructed.
 It is this structure for the wave function $\psi(x,t) \equiv
 <x,t|\psi)$, where $x$ and $t$ are the spectra of self-adjoint
 operators, that provides the possibility of coherence in $t$, and
 therefore, interference phenomena. If the quantum theory is to remain 
symplectic in form, the variable $E$ must also be adjoined.
\par  The standard nonrelativistic quantum theory cannot be used to
 predict interference in time.  For example, Ludwig$^3$ has pointed out
that the time variable cannot be a quantum observable, since there is no
imprimitivity system (i.e., no operator exists that does not commute with
$t$ in the nonrelativistic theory) involving this variable. Note that
the Hamiltonian of the standard theory evolves quantum states in time,
but does not act as a shift operator since it commutes with $t$.Dirac$^2$
has argued that if $t$ were an operator, then the resulting
$t,E$ commutation relation would imply that the energy of the system
is unbounded below, from which he concluded that the time cannot be an
observable in the nonrelativistic quantum theory (note, however, that
in a relativistic theory, negative energies correspond to
antiparticle states, and are not excluded).  Moreover, as the
axiomatic treatment of Piron$^4$
(see also, Jauch$^5$) shows, the Hilbert space of the quantum theory is
constructed of a set of wave functions satisfying a normalization
condition based on integration over all space,  e.g., for a single
particle, $\int \vert \psi_t(x)\vert^2 d^3x
\leq \infty$, for each value of the parameter $t$.  There is therefore
a  distinct Hilbert space for each value of the parameter $t$.
  A simple argument based on the propagator
for the Schr\"odinger equation demonstrates that interference in
time cannot occur in the standard Schr\"odinger theory.
\par  The propagator for wave functions in the standard Schr\"odinger
 treatment, is given by
$$\eqalign{\langle x| U(t)| x'\rangle &=  \bigl({m\over 2\pi
it}\bigr)^{3\over 2} e^{i{m\over 2t}( x- x')^2}\cr
              &= G( x- x',t),\cr} \eqno(1)$$
where $x$ is here a three dimensional variable.
The action of the propagator is  
 $$ \psi_t(x) = \int
dx'G(x-x',t-t')\psi_{t'}(x').\eqno(2)$$
The integration over $x$ makes possible the description of the double
slit experiment in space (by coherently adding up contributions from
two or more locations in $x$ at a given $t'$); there is, however, no
integration over $t'$, and therefore no mechanism for constructing
 interference in time. This result, obvious from the form of
$(2)$, is a reflection of the arguments of Ludwig$^3$ cited above,
and is fundamental to the standard nonrelativistic quantum theory.
This structure constitutes a formal proof that no interference effect
in time is predicted by the standard nonrelativistic Schr\"odinger
theory.  Introducing two packets into the beam of an experiment at two
different times $t_1$ and $t_2$ would result in the direct sum of the
two packets at some later time, say $t_3$, if one propagates the first
from $t_1$ to $t_3$, and the second from $t_2$ to $t_3$.  This would
constitute, by construction, a mixed state, for which no interference
would take place, just as the construction of a beam of $n+m$
particles by adding a set of $n$ particles with definite
spin up to another set of $m$ particles with spin down results in a
mixed state descrobed by a
diagonal density matrix with {\it a priori} probability $n/(n+m)$ for
outcome spin up, and $m/(n+m)$ for outcome spin down.  There is no
coherent superposition which would result in some spin with certainty
in {\it any} direction$^5$.
\par Moreover, as
pointed out by Wick, Wightmann and Wigner$^6$, a Hilbert space
decomposes into incoherent sectors if there is no
observable that connects these sectors; hence, if there were a larger
Hilbert space containing a representation for $t$, the absence of any
observable that connects different values of $t$ in the standard 
nonrelativistic
physics would induce a decomposition of the  the Hilbert space into a 
(continuous) direct
sum of superselection sectors$^4$. Therefore, no superposition of vectors for
different values of $t$ would be admissible. This would exclude the
 interpretation
of the experiment given in ref.1  forming the basis of the analysis
carried out by the authors involving the linear superposition of two
parts of a particle
wave function arriving at the detector simultaneously, but
originating at two different times, in the framework of the standard
 nonrelativistic quantum theory.
\par The significance of the experiment of ref. 1 is 
that it demonstrates at least one class of phenomena actually seen to occur in
nature at low energies (but high frequency),  for which the standard
 nonrelativistic quantum theory does not provide an adequate description,. 
and this fact demands the
development of some new theoretical tools which are a proper
generalization of the standard theory.
\par  The situation for particles, in this respect,
is very different from that of electromagnetic waves, for  which the
second order equations imply coherence in time as well as space (the
coherence time for light waves is a commonly measured characteristic
of light sources).  It is clear from the (spatial) double slit interference of
light, which travels at a fixed velocity, that the sections of a wave
front passing through the two slits must pass at different times if
they are to arrive simultaneously at the detection plane
off-center. The arrival of pieces of a {\it particle} wave packet which have
passed through two spatially separated slits simultaneously on a
screen off center is made possible by the dispersion of momenta in the
wave packet, permitting a range of velocities.  If the two
contributions to the linear superposition on the screen were not taken
to  be 
simultaneous at the two slits, they would not interfere, since
they would have originated on wave packets at different values of
time.  If, indeed, such interference could take place, we would have to
add up the contributions passing each slit for all times, and this
would destroy the interference pattern (one can see in the standard
calculation in every textbook that the two pieces of the wave packet
that contribute to the interference are taken at equal time, i.e.,
from a single wave packet arriving at the slits). Although this
example is given in every quantum mechanics text as an essential
property of the quantum theory, the possibility of interference in
time is never mentioned.
  \par Moshinsky$^7$, in 1952, raised the question of interference in
time.  His calculation, however, was concerned with the evolution of a
single wave packet, passed through a spatial slit opened at time
$t=0$.  The transient form of the wave function was then calculated;
it has the appearance of a Fresnel interference pattern. Using
semiclassical time of flight arguments, it was deduced that this
behavior could be thought of as an interference in time. The actual 
superposition of wave functions at two different times was not
considered.
\par If we were to assume, arbitrarily, that the waves from sources at
two different times could be coherently added (note that Eq.(1) is
only valid for the evolution of physical states up to an undetermined
c-number phase), the propagator above
could be expanded in a power series for small variations in the final
time, and one would find some semblance of an interference pattern for
a few maxima before distortion would set in. For $\varepsilon$ the
time between peaks of the laser beam, $T$ the time between peaks on
the predicted ``interference pattern'' on a screen at a distance 
$L$ from the emission source, $m$ the mass of the electron, one finds
a crude estimate 
$$ \varepsilon T \sim {\pi \hbar\over 2} \sqrt{{m \over 2}}L(E^e_{kin})^{-{3
\over 2}},$$
where $E^e_{kin}$ is the kinetic energy of the emitted
electron, and $L$ is the distance from emitter to the detector.
 Sali\`eres, {\it et al\/}$^8$, remark that in
the very nonlinear emission of an electron by a high energy laser
beam, several hundred photons may be absorbed to be converted to
kinetic energy. If we take the electron kinetic energy as equal to the
energy $\hbar \omega$ of the laser beam, for this experiment,
approximately $1.46 eV$, with a factor of 300, and $L\sim 1 cm$, we 
obtain for the
first several predicted peaks (before distortion due to nonconvergence of
the power series expansion) a value of $\varepsilon T \sim 9
\times 10^{-28} sec^2$, or, for  $\varepsilon \cong T$, a diffraction
spacing of $T \sim 3 \times 10^{-14}= 30 fs$.  This calculation  indicates
how an exact solution
of the time dependent Schr\"odinger equation could exhibit an interference-like
pattern for several peaks on the detector plane in agreement with the
experimental results.. The numerical
predictions of this integration do not,
however, provide reliable evidence that the procedure is consistent
with the theoretical basis of the standard nonrelativistic quantum theory.
\par There appear to be several types of theories which accommodate
 time as an observable. These are, in a nonrelativistic
framework,  Floquet theory$^9$, for which the evolution operator has
the form $E+H$, where $H$ is a standard Hamiltonian model (often $t$
dependent) and $E$
corresponds to the operator $i\partial/\partial t$, a  formulation of quantum
theory in which
one of the space variables becomes the evolution parameter and the
time variable becomes an observable$^{10}$, the quantum Lax-Phillips
theory for irreversible processes (also applicable to relativistic
quantum theory)$^{11}$, and in a relativistic framework, the so-called
constraint theories$^{12}$ in addition to the theory developed by 
Stueckelberg$^{13}$.  In the
book edited by J.G. Muga, {\it et al}$^{14}$, much discussion is devoted to the
difficulties of treating time as an operator in the standard
framework, and some constructions are given for which certain operator
valued functions of the phase space variables $q,p$ can be interpreted
as ``time operators''.The spectra of these operators do not, however,
enter into a propagator in a way that can naturally generate
interference effects of the type observed in Lindner, {\it et al}.
 \par A time operator, corresponding to the variable $t$ that we
 recognize as time in the laboratory, is constructed in the work of 
Hahne$^{10}$, in which he permits a space coordinate, say, $z$ to
act as an evolution parameter, and $x,y,t$ become operator valued.
 Although coherence in $t$ can be
achieved, the use of $z$ as an evolution parameter for the flow of
the wave packet makes it difficult to formulate a description of the
Lindner,{\it et al}, experiment in these terms.
  \par The quantum Lax-Phillips theory$^{11}$ provides a systematic and
rigorous description of irreversible processes.  In this theory, a
 unitary evolution by a parameter, say $s$, is introduced on a Hilbert
 space ${\bar H}$, which is foliated along the spectrum $t$ of a
 ``time'' variable
 which is a self-adjoint operator on  ${\bar H}$ into a set of
 Hilbert spaces $H_t$, which may be identified with the Hilbert spaces
 of the ordinary quantum theory, but maintains its coherence in
$t$. The existence of invariant subspaces delimited by time intervals
makes it possible to construct semigroups in these subspaces.
The simplest model for evolution in the nonrelativistic case is of the
 form $E+H$, coinciding, as we shall see below, with the Floquet
 construction.  Although there are applications of Lax-Phillips theory
 to the description of the experiment of Lindner,{\it et al}, the
 discussion of the Floquet construction below will be adequate for
 our present purposes.
\par The constraint theories$^{12}$, primarily directed toward a
description of relativistic dynamics, are constructed from a set of
 functions $K_i$, where $i=1,2,...N$ for a system of $N$ particles;
 each of the form $p_i^\mu {p_i}_\mu +m_i^2 + \phi_i$, where the
 interaction terms $\phi$ are functions of all the momenta and
 coordinates of all of the particles (here, $\mu = 0,1,2,3$).  In the
 classical form of the theory, the $K_i$'s are taken to be zero on a
 constraint hypersurface, and satisfy the requirement that they be
 first class contraints (vanishing Poisson brackets).  The evolution
 of the system goes with some parameter, say $s$, according to the
 transformations induced in the phase space by some linear combination
 of the $K_i$'s. For the two particle case, for $\phi$ of the form
 $\phi_1(x_1-x_2)=\phi_2(x_2 - x_1)$, it is easy to see that this
 theory is equivalent to the Stueckelberg theory$^{13}$ we shall discuss
 below (the relative motion for which $x = x_1-x_2$ appears as
 a one-body problem).  We therefore will not discuss the constraint
 theory further here.  In following, I discuss the Floquet theory and
 the Stueckelberg theory as somewhat categorical among the set of
 theories which may be used to describe the Lindner, {\it et al},
 experiment. Furthermore, since, as we shall point out below, the
 nonrelativistic limit of the Stueckelberg theory goes over to the
 Floquet form, it appears as the main candidate for a viable
 description.  Although the Stueckelberg theory is essentially
 relativistic, and the energies of the macroscopic motions of the
 particles involved are low, the very high frequencies used to
 establish excitations and pulse rates involve high energy components
 of the wave packets, and thus the use of a relativistically covariant
 theory is appropriate.
 \par Floquet theory$^9$ was originally intended for the treatment of 
differential equations with periodic coefficients.  It entered
physics in an important way in solid state theory where the potential
 in a crystal is
periodic in space.  Utilizing the translation operator 
$ U({\bf a}) = e^{i{\bf p}\cdot {\bf a}},$
where ${\bf a}$ is a crystal lattice vector,
one can show that the solutions of the Schr\"odinger equation, as a
representation of this translation group, take on the Bloch form.
\par The idea then arose that for a Hamiltonian periodic in time, the
same method could be used.  However, since the Hamiltonian commutes
with  $t$, to make the group action explicit, it was necessary to introduce
a new variable $E$ (the generator of translations in $t$).  The  
 evolution operator was then defined as
$$ K= E+H,$$
where $E\equiv -i\hbar \partial_t$.  Then, clearly,for 
$$U(T) = e^{-iKT} $$
the operator $U(T)$ carries $t\rightarrow t-T $, translating functions
of  $t$ to the right by $T$. 
\par  The introduction of this modification of
the Hamiltonian was also suggested by Howland$^{15}$ for both the classical and
quantum theories for treating problems in which the Hamiltonian depends
on time.  For the classical theory,
introducing a new parameter of evolution, say $s$, the Hamilton
equations would then include the relations
$$ {dt \over ds } = {\partial K\over \partial E}$$
and 
$${dE \over ds} = -{\partial K \over \partial t} = -{\partial H \over
  \partial t}, $$
thus providing some interpretation for $E$.  Since then ${dt \over ds} = 1,$
by a change of variables, this formulation becomes completely
equivalent to the standard form.  However, in the quantum theory, the
Hamilton equations, as operator equations, imply conditions on
expectation values; the variables $t$ and $s$ are then no longer
equivalent. In this case, $s$ is the parameter of the motion, and $t$
is a quantum operator, an observable.  The wave functions are then
coherent in $t$, making possible, in principle, interference phenomena in $t$.
\par The resulting theory is very different from the
standard Schr\"odinger theory.  To see this, let us write the
corresponding evolution equation in what I shall call ``Floquet theory'',
since it has the same structure for the Hilbert space,
but I will not insist that $H$ be periodic in $t$.  The
mathematical framework is independent of this periodicity (clearly the
consequences of the theory, and the results one may obtain, can be very strong
 when $H$ is periodic in $t$).
\par The evolution equation has the form
$$\eqalign{ i{\partial \psi_s \over \partial s} &= K\psi_s\cr
                       &= (-i\hbar \partial_t + H)\psi_s \cr} \eqno(3)$$
where $\psi_s$ is a function of $x,t$. The functions $\psi_s$ have the
property that
$$  \Vert \psi_s \Vert^2 =\int d^3x dt \vert \psi_s( x,t)\vert^2
  <\infty, \eqno(4) $$
the condition that $\psi_s$ belongs to a Hilbert space $ H_s$
  (now labelled by $s$).
  As pointed out by Kulander and
Lewenstein$^{16}$, if $H$ (or $K$) is periodic over some hundreds of cycles,
it would be  a good approximation to assume a ``stationary state'' in
which the  $s$ derivative of $\psi_s$ is replaced by an eigenvalue
(their equation (72.31)). Such a state would be stationary in $s$, not
  $t$; the idea is that the spacetime function $\psi_s(x,t)$ reaches a
  steady form and no longer changes, on the spacetime manifold, as a
  function of $s$ (up to a phase determined by the eigenvalue). In
  this case, the solution  of eq.(3) amounts formally to an
  integration of the time dependent Schr\"odinger equation over
  $t$, as carried out in the analysis of ref. 1,
 with the Hamiltonian shifted by
  the Floquet eigenvalue (possibly zero).  The theory would then
  predict coherence in $t$ for such a solution. This is not, however,
  a valid procedure for the conditions of the experiment of ref. 1,
  since this experiment involves essentially just a couple of cycles.
\par In the following I calculate the propagator
for the Floquet equation (3) for the case of a {\it free} particle. I will
 show that even though interference in
$t$ is, in principle, possible in the Floquet framework, two narrow
segments, in time, of a particle wave function will not interfere
 unless
 (a) the segments initially overlap, or (b) there is a nontrivial 
$t$-dependence
(but not necessarily periodic) in $H$, the latter certainly providing
an interesting possibility for application to the experiment we are
 discussing.  
\par To obtain the form of the propagator, let us consider the $x,t$
matrix elements of the unitary evolution $U(s)$ of  $\psi_s$:
$$\langle x,t| U(s)|x',t'\rangle = \int dE'dEdp'dp''
\langle x,t|E'p'\rangle \langle E'p'|e^{-i(H-E)s} |E''p''\rangle 
\langle E''p''|x't'\rangle  \eqno(5)$$
Here, the momenta and coordinates are three dimensional (the
differentials are also $dp\equiv d^3p$). 
\par I now assume that $H$ has the free particle form $p^2/2m$  and
therefore commutes
with $E$. Then, $(5)$ becomes 
$$\langle x,t| U(s)|x',t'\rangle= \delta(t'-t+s) \bigl({m\over 2\pi
is}\bigr)^{3\over 2} e^{i{m\over 2s}(x-x')^2}. \eqno(6)$$
Let us now call the coefficient of the $\delta$ function $G(x-x',s)$.
The propagation of $\psi_{s'}(x'.t')$ to $\psi_s(x,t)$ is given by
$$ \psi_s(x,t) = \int
dx'dt'\delta(t'-t+(s-s'))G(x-x',s-s')\psi_{s'}(x',t')\eqno(7),$$
 clearly displaying the possibility of interference in $t$, i.e.,
there may be contributions at several different
values of $t'$ corresponding to the opening of time gates. However, there is
no spreading, in this propagation, of the width of the time pulses,
independently of the form of the wave packet $\psi_{s'}(x',t')$.
\par Consider the contribution of two gates at $t_1$ and $t_1'$.  In
this case,
$$\eqalign{\psi_s(x,t) &= \int dx' G(x-x',s-s') \cr
&\{ \delta((s-s') - (t-t_1)) \psi_{s'}(x',t_1)\Delta t_1
+ \delta((s-s') - (t-t_1'))\psi_{s'}(x',t_1')\Delta t_1'\},
\cr}\eqno(8)$$
where $\Delta t_1$ and $\Delta t_1'$ are the (narrow) widths of the gates.
 \par One can evaluate  $s-s'$ approximately through the Hamilton
equations. Since $dx/ds = p/M$, it
follows that $\Delta s \cong ML/p,$ where $L$ is the distance from
source to detector.  Since $d<t>/ds =1$, the expectation value of $t(s)$
goes with
$s$, so that $s-s' \cong t-t'$, the latter giving the time from the
source gate to the time on the detector when the measurement is made.
 Due to the delta function constraint,
 we see that there can be no interference if the source
pulses do not overlap. Alternatively, the delta functions would not
appear if the
Hamiltonian had an explicit $t$-dependence, and the result would
depend on the particular model. Interference in the
framework of the Floquet structure, therefore, although in principle
possible, would not occur for narrow source pulses in the absence of
explicit time dependence in the Hamiltonian.
\par The  Floquet theory is, in fact, a
nonrelativistic limit of Stueckelberg's relativistic quantum
theory$^{13}$, the second way of introducing coherence in time which I
wish to discuss here.  In 1976, Horwitz and
Rabin$^{17}$ pointed out that the relativistic quantum theory of Stueckelberg
 predicts inteference in time.  In this theory, $t$ is treated as a
quantum observable, since the Einstein variables $x,t$ are considered,
in relativity, as the nontrivial outcome of experiments measuring the
place and time of occurrence of events.  Their calculation will be
 briefly redone below for the parameters of the experiment of ref.1.
In this theory, interference does not require initial overlap
or an explicitly time dependent Hamiltonian. The estimate given below
shows that the interference criteria are satisfied with numbers very
close to the conditions and results of the experiment under
discussion; the high frequencies required
are due in this case to the large value of the velocity of light.
\par The Stueckelberg theory for the free particle introduces an
equation quite similar to that of the Floquet equation, but with an
evolution operator that is Lorentz invariant:
$$ i\hbar {\partial \psi_s\over \partial s} =
 { p^2-({E\over c})^2 \over
2M}\psi_s, \eqno(9)$$
where $\psi_s(x,t)$ satisfies the same normalization condition as for
the Floquet theory, on space and time,i.e., $\int \vert
\psi_s(x,t)\vert^2 d^4x  \leq \infty$, and $M$ is the Galilean target
mass (the so-called mass shell value for $m^2c^2=(E/c)^2 - {\bf p}^2$).
  The propagator has a similar form to that of the Floquet propagator,
but is Gaussian in all four variables:
$$\langle x| U(s)|x'\rangle=  \bigl({M\over 2\pi
is\hbar}\bigr)^2 e^{i{M\over 2s\hbar}(x-x')^2}, \eqno(10)$$
where now $(x-x')^2$ is the invariant $({\bf x}-{\bf x'})^2 -
c^2(t-t')^2$; we write $x$ for $(x,t)$. It is the quadratic term in $t-t'$ in
 the exponent which
leads directly to interference in the same way as the double slit in
space.
 \par   The diffraction formula,
obtained from $(10)$,  using the
Hamilton relations
$$ {dx\over ds} = {p \over M },\eqno(11)$$
and  (note that this relation allows for two pulses emitted at
different times to arrive at a detector at the same time due to the
spread in the spectrum of $E$)
 $$ {dt \over ds} = {E \over Mc^2}.\eqno(12)$$
is$^{17}$
$$  \varepsilon T \cong {2\pi \hbar L \over <p>c^2}, \eqno(13)$$
where $\varepsilon$ is the gate spacing in time, and $T$ is the time
between diffraction peaks at a distance $L$.
\par For 850 nm light, as utilized in the experiment under discussion, as
remarked above, 
$\hbar \omega$ is about $1.46 eV$. Using the on-shell
value for the electron mass, 
taking into account (as assumed above) that the
electron may absorb about $300$ photons during the emission, $cp$ (for
$p$ in the beam direction)
then has a value of $1.21 \times 10^3 eV$.
  With these values, one finds that ,
$$ \varepsilon T \cong 6.9 \times 10^{-30} sec^2, \eqno(14)$$
so that for $\varepsilon\sim T$, $T\sim 2.6 \times 10^{-15} sec$. This
result, for the pulse rate and the observed diffraction pattern, is in good
agreement with the results obtained in the experiment.
\par More precise estimates can be obtained by taking into account
more details of the interaction, and the dependence on $L$ can be used
 as a parameter to test the reliability of $(13)$.
\par  The relativistic model
therefore seems to provide a simple and consistent description of the
experimental results. At very low energies, the
Stueckelberg theory, which carries a clear interpretation of $t$ as an
observable, reduces approximately to the Floquet form$^{18}$. The estimate
made above therefore includes the result that would be obtained
 approaching the
low energy Floquet type as a limiting case.
\par I thank C. Piron for communications on the question of
coherence, and Gerhard Paulus, one of the authors of ref. 1, for
discussions of this experiment and the analyses applied. I would also
like to thank Igal Aharonovich for bringing the initial announcement
of this experiment to my attention and Noam Erez and Jonathan Silman
for useful discussions.
\bigskip
\noindent {\bf References}
\frenchspacing
\item{1.} F. Lindner, M.G. Sch\"atzel, H. Walther, A.Baltuska,
E. Goulielmakis, F. Krausz, D.B. Milo\v sevi\'c, D. Bauer, W. Becker
and G.G. Paulus, quant-ph/0503165, Phys. Rev. Lett.{\bf 95}, 040401 (2005).
See also,Pascal Szriftgiser, David Gu\'ery-Odelin,Markus Arndt and
Jean Dalibard, Phys. Rev. Lett. {\bf 77} 0031-9007 (1996);
M. Wollenhaupt, A Assion, D. Liese, Ch. Sarpe-Tudoran, T. Baumert,
S. Zamith, M.A. Bouchene, B. Girard, A Flettner, U. Weichmann and
G. Gerber, Phys. Rev. Lett.{\bf 89},0031-9007(2002).
\item{2.} P.A.M. Dirac,{\it Quantum Mechanics}, First edition,pp.34,36, Oxford
Univ. Press, London (1930); {\it Lectures on Quantum Field Theory},
Academic Press, New York (1966).  See also W. Pauli, General
{\it Principles of Quantum Mechanics},p.63, Springer-Verlag, Heidelberg
(1980);E. Schr\"odinger Berl. Ber., p. 238 (1931); P. Carruthers and
M.M. Nieto, Rev. Mod. Phys. {\bf 40},411 (1968), and references
therein.
\item{3.} G. Ludwig,{\it  Foundations of Quantum Mechanics I}, p.295,
Springer-Verlag, New York (1983).
\item{4.} C.P. Piron, {\it Foundations of Quantum Physics}, W.A. Benjamin,
Reading (1976). See also, C. Piron, physics/0204083 29 Apr. 2002; {\it
Trends in Quantum Mechanics,} eds. H.-D. Doebner {\it et al},p.270,
World Scientific, Singapore (2000).
\item{5.}J.M. Jauch, {\it Foundations of Quantum Mechanics}, Addison-Wesley,
Reading (1968).
\item{6.} G.C. Wick, A.S. Wightman and E. Wigner, Phys. Rev. {\bf
88}, 101 (1952).  See also  C. Piron,
Helv. Phys. Acta {\bf 42}, 330 (1969).
\item{7.} M. Moshinsky, Phys. Rev. {\bf 88} 625(1952).
\item{8.} P. Sali\`eres, B. Carr
'e, L. Le D
'eroff, F. Grasbon, G.G. Paulus, H. Walther, R. Kopold, W. Becker,
D.B. Milosevic, A. Sanpera, M. Lewenstein, Science {\bf 292}, 902(2001).
\item{9.}For example, H.L. Cycon, R.G. Froese, W. Kirsch and B. Simon,
{\it Schr\"odinger Operators with Application to Quantum Mechanics and
Global Symmetry}, p. 146, Springer-Verlag, New York (1987); see also,
R.M. Potvliege and R. Shakeshaft, {\it Atoms in Intense Laser Fields},
ed. M. Gavrilla, Academic Press, San Diego (1992); S.I. Chu,
Adv. Chem. Phys. {\bf 73},739 (1989). Note that the definition of
$H-i\hbar\partial_t$ as a self-adjoint operator is sufficient to
define the structure of the Hilbert space and imply Eq. $(5)$.
\item{10.} G.E. Hahne, Jour. Phys.A{\bf 367}, 144 (2003)
\item{11.} Y. Strauss, L.P. Horwitz and E. Eisenberg, Jour. Math. Phys. {\bf
41}8050(2000), Fermilab-Pub-97/227-T. The original form of the the
theory, intended for application to classical wave equations such as for
optical or acoustic waves, is worked out in the book {\it Scattering
Theory}, P.D. Lax and R.S. Phillips, Academic Press, New York
(1967). See also, L.P. Horwitz and C. Piron, Helv. Phys. Acta {\bf
66}, 693 (1993). 
\item{12.} F. Rohrlich, Phys. Rev. D{\bf 23}, 1305 (1980), and
references therein to earlier work.
\item{13.}  E.C.G. Stueckelberg, Hel. Phys. Acta {\bf 14}, 372,585
(1941); {\bf 15}, 23 (1942); see also,  L.P. Horwitz and C. Piron,
Helv. Phys. Acta {\bf 46}, 316 (1973).
{\it Proceedings of the Workshop on Constraints Theory and Relativistic
Dynamics,} Arcetri, Firenze, Italy, 1986, ed. G. Longhi and L. Lusanna,
World Scientific, Singapore (1987);
L.P. Horwitz and F. Rohrlich, Phys. Rev. D{\bf24}, 1528 (1984);
K. Sundermeyer, {\it Constraint Dynamics}, Lecture Notes in
Physics,vol.169, Springer, Berlin (1982).
\item{14.} {\it Time in Quantum Mechanics}, ed. J.Muga, R. Sala Mayato
and I.L. Egusquiza, Lecture Notes in Physics, Springer, Berlin (2002).
\item{15.} J.S. Howland, Indiana Math. Jour. {\bf 28}, 471 (1979). 
\item{16.} K.C. Kulander and M. Lewenstein, {\it Atomic, Molecular and
Optical Physics Handbook} (G.W. Drake, ed.), p. 828, American Institute of
Physics Press, Woodbury, N.Y. (1996).
\item{17.} L.P. Horwitz and Y. Rabin, Lett. Nuovo Cimento {\bf 17},
501 (1976).
\item{18.}  L.P. Horwitz, W.C. Schieve and C. Piron, Ann. Phys. {\bf
137}, 306(1981); L.P. Horwitz and F. Rotbart, Phys. Rev. D {\bf 24}, 2127
 (1981). 

\end
\bye